\documentclass{kluwer}    
\usepackage[dvips]{epsfig}
\newdisplay{guess}{Conjecture}

\begin{document}                                                                                   
\begin{article}
\begin{opening}         
\title{Searching for an Intrinsic Stellar Population in 
       Compact High-Velocity Clouds
       \thanks{Based on observations obtained at the ESO-VLT, Paranal, run
        67.B-0060(A)}} 
\author{Ulrich \surname{Hopp}\email{hopp@usm.uni-muenchen.de}}  
\runningauthor{U. Hopp, R.E. Schulte-Ladbeck, J. Kerp}
\runningtitle{Stellar Content of CHVC}
\institute{Universit\"ats-Sternwarte M\"unchen, Scheiner-Str.1, D 81679
            Munich, Germany}
\author{Regina E. \surname{Schulte-Ladbeck}\email{rsl@phyast.pitt.edu}} 
\institute{University of Pittsburgh, Pittsburgh, PA 15260, USA}
\author{J\"urgen \surname{Kerp}\email{jkerp@astro.uni-bonn.de}} 
\institute{Radioastronomisches Institut der Universit\"at Bonn, 
           Auf dem H\"ugel 71, Bonn, Germany}

\begin{abstract}
We are investigating the hypothesis that Compact High--Velocity Clouds (CHVCs)
are the left-over building blocks of Local Group galaxies.
To this end, we are searching for their embedded stellar populations using
FORS at the VLT. The search is done with single-star photometry in V and I bands, 
which is sensitive to both, young and old, stellar populations.
Five CHVCs of our sample have been observed so far down to I=24. 
We pointed the VLT towards the highest HI column density regions,
as determined in Effelsberg radio data. 
In an alternate approach, we searched 2MASS public data towards those 5 CHVCs down to K=16. While 
the VLT data probe the central regions out to distance moduli of about 27, the 2MASS data are
sensitive to a population of red giant stars to distance 
moduli of about 20. The 2MASS data, on the other hand, cover a much wider field of view than the
VLT data (radius of 1 degree versus FORS field of 6.8 arcmin). 
We did not find a stellar population intrinsic to the
CHVCs in either data. In this paper,  we illustrate our search methods.
\end{abstract}
\end{opening}           

\section{Introduction}

Recent cold dark matter simulations of the formation and evolution
of galaxies predict the existence of a significantly higher amount
of substructure around big galaxies like the Milky Way \cite{KLY}, \cite{MOO} 
than observed in the form of dwarf galaxies \cite{MAT}. One solution for
this so-called dwarf galaxy crisis could be that the predicted
subhalos have been overlooked observationally and are hidden among
the population of Compact High-Velocity Clouds (CHVC). Blitz et al. (1999) 
suggested that isolated CHVCs might be the leftover building blocks
predicted in the CMD scenario with mean distance of about 1~Mpc.
Braun \& Burton (1999, 2000) identified an intial catalog of 65 CHVCs.
We here present deep optical VLT imaging and 2MASS archival studies
of five CHVCs to test them for the presence of a stellar
population.

\begin{figure}[] %
\centerline{
\epsfig{file=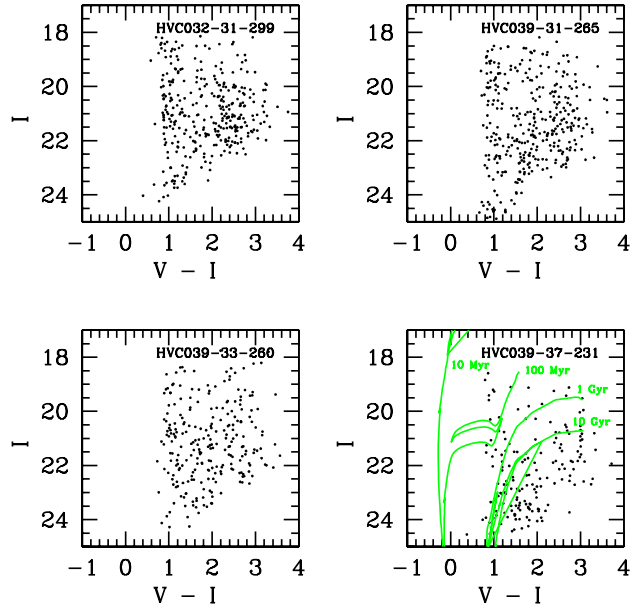,width=0.8\textwidth}
}
\caption[]{I, V-I CMDs of 4 CHVC fields. The stellar populations 
in all fields resemble a galactic, high latitude field population. There is
no trace of either a young, or an old stellar population intrinsic to any of the CHVCs. In the lower right panel, Giradi et al. (2000) 
isochrones (1/5 solar) are overplotted for 
10~Myr, 100~Myr, 1~Gyr, 10~Gyr and a distance of 1~Mpc to indicate
the expected range of the intrinsic stellar populations.
}
\label{hoppu.FIG1}
\end{figure}

\begin{figure}[] %
\centerline{
\epsfig{file=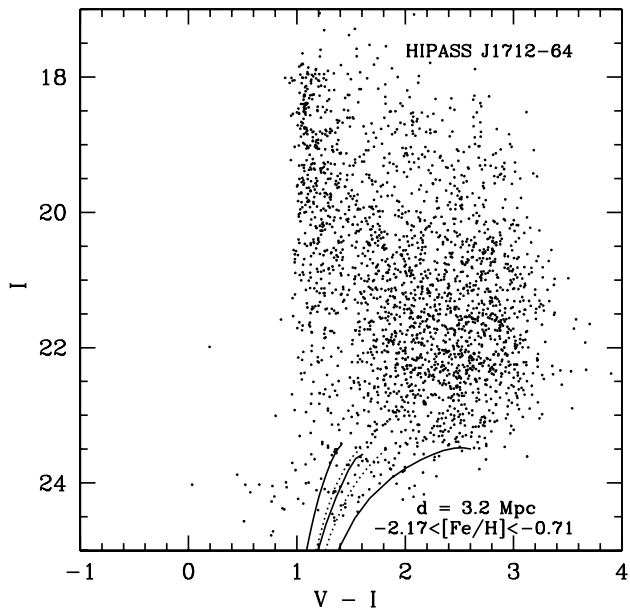,width=0.8\textwidth}
}
\caption[]{I, V-I CMD of the field of HIPASS~J1712-64. 
Globular cluster ridge-lines of Da Costa \& Armandroff (1990)
are overplotted.
}
\label{hoppu.FIG1}
\end{figure}

\section{Observations}

V and I band observations were obtained with FORS1 at the VLT down
to limiting magnitudes to be sensitive to red giant stars at distances
up to $\sim$~3~Mpc. The field of view, 6.8 by 6.8 arcmin, was 
centered on the highest column density regions (for four CHVCs) using Effelsberg radio data. 
We also selected all stars within a radius of 1$^o$ 
from 2MASS (second incremental data release). These near-infrared JHK data are sensitive
to old stellar populations at the typical distances of Milky Way
dwarf Spheroidal companions. 

The optical CCD data were searched and analyzed with DAOPHOT~II \cite{STE}.
They are not only sensitive to old stars but would easily also reveal
the presence of main sequence (MS) stars of a young stellar population 
to large distances.

\section{Results}

Fig.~1 shows color magnitude diagrams (CMD) for
four of our CHVC fields. Fig.~2 is the CMD of a HIPASS HVC. Extended sources
were excluded from the analysis; the amount of unresolved background galaxies 
is small anyway due to the excellent seeing of the data, namely
0.65 arc sec FWHM. The CMDs appear as expected for Milky Way
halo fields. A visual inspection does not reveal any additional
stellar component, e.g. a RGB or a MS at the expected distances/apparent magnitudes. 

A visual inspection of near-infrared 2MASS CMDs does not show any evidence 
for a spread-out RGB or AGB
population. For comparison, 2MASS CMDs of several dSph Milky Way
companions, as well as positional plots of the stars, allow the detection of intermediate 
rich systems like For and Scl, while UMi can not be isolated from the halo stars.

We performed statistical tests by adding artificially, diluted red giant branches of 
known dwarf galaxies, to the observed CHVC CMDs. We then compared
these distributions to those originally observed using a
Kolmogorow-Smirnow test. A population of about 50~RGB stars at a distance of
$\sim$~0.5~Mpc is at the detection limit of the VLT data; richer systems such as
those of the known dSph Milky Way satellites would easily be found.
For details of this analysis, see Hopp, Schulte-Ladbeck \& Kerp, 2003
(MNRAS, in press).

\section{Conclusions}
We conclude that the observed compact clouds do not host an 
intrinsic stellar population. 
Our conclusions agree with and extend those reported by 
Simon \& Blitz (2002), who did not detect
stars in CHVCs on processed POSS scans.

\begin{acknowledgements}
This research has made use of the NASA/IPAC Infrared Science
Archive, which is operated by the Jet Propulsion Laboratory,
California Institute of Technology, under contract with the National
Aeronautics and Space Administration. UH acknowledges support 
by the SFB 375 of the DFG. RSL thanks the Department of Physics \&
Astronomy for a leave of absence, and the MPE Garching for hosting her visit.
\end{acknowledgements}

\end{article}
\end{document}